\documentclass[12pt]{article}
\usepackage{float}
\usepackage{amsmath}
\usepackage{amssymb}
\usepackage{amsfonts}
\usepackage{latexsym}
\usepackage{color}
\usepackage{graphicx}
\usepackage{verbatim}
\usepackage{subcaption}
\catcode `\@=11 \@addtoreset{equation}{section}
\def\theequation{\arabic{section}.\arabic{equation}}
\catcode `\@=12

\newcommand{\be}{\begin{equation}}
	\newcommand{\en}{\end{equation}}
\newcommand{\bea}{\begin{eqnarray}}
	\newcommand{\ena}{\end{eqnarray}}
\newcommand{\beano}{\begin{eqnarray*}}
	\newcommand{\enano}{\end{eqnarray*}}
\newcommand{\bee}{\begin{enumerate}}
	\newcommand{\ene}{\end{enumerate}}

\newcommand{\B}{{\mathfrak B}}

\newcommand{\Cf}{{\mathfrak C}}
\newcommand{\Ff}{{\mathfrak F}}

\newcommand{\mc}{\mathcal}
\newcommand{\norm}[1]{ \parallel #1 \parallel}
\newcommand{\ch}{{\hat C}}
\newcommand{\fh}{{\hat F}}
\newcommand{\rh}{{\hat R}}

\newcommand{\Sc}{{\cal S}}

\newcommand{\G}{{\cal G}}

\newcommand{\C}{{\cal C}}
\newcommand{\1}{1 \!\! 1}

\newcommand{\Hil}{\mc H}

\renewcommand{\l}{\langle}
\renewcommand{\r}{\rangle}

\newcommand{\pin}[2]{\l#1 , #2\r}

\newtheorem{thm}{Theorem}

\newtheorem{prop}[thm]{Proposition}

\catcode `\@=11 \@addtoreset{equation}{section}
\catcode `\@=12

\begin{document}

\thispagestyle{empty}

\vspace*{1.6cm}

\begin{center}
{\Large \bf A quantum-like  model of political consensus via non self-adjoint Hamiltonians}   \vspace{2cm}\\

{\large F. Bagarello}\\
Dipartimento di Ingegneria,
Universit\`a di Palermo,\\ I-90128  Palermo, Italy\\
and I.N.F.N., Sezione di Catania\\
e-mail: fabio.bagarello@unipa.it\\

\vspace{5mm}

{\large G. Liarda}\\
Dipartimento di Fisica e Chimica,\\ Universit\`a di Palermo,\\ Via Archirafi 36,\\ I--90128  Palermo, Italy\\
e-mail: gloria.liarda00@gmail.com\\

\end{center}

\vspace*{1cm}

\begin{abstract}
	We discuss here how non self-adjoint Hamiltonians, and their related Heisenberg-like dynamics, can be used to model a political system consisting in a coalition $\C$ of different parties (forming a government) and by their (original) supporters $\Sc$. Our aim is to model how the opinion of these supporters changes depending on the efficiency, competence and coherence of the coalition $\C$, as these are perceived by $\Sc$ during their action while governing. After a rather general introduction we propose three specific models, and we describe and comment  the dynamical behaviour of the { full} system, $\Sc\cup\C$.  The role of the so-called {\em balanced Hamiltonians}, recently introduced by the authors in connection with integrals of motion, is discussed in details.

\end{abstract}

\vspace{2cm}


\vfill


\newpage

\section{Introduction}

A rather common feature we observe in politics, not only in our Country, is that there are serious differences between what is stated by those people trying to be elected and what they concretely do {\bf after} being elected. Coherence is often considered not really a value by many politicians, especially by those who were part of the minority {\bf before} the elections. Quite often these people, once elected, have to face with  real life, and start realizing that, if they really want to avoid a default for the Country they now lead, it is not possible, for instance, lower the taxes too much, allow people to retire sooner, give stronger financial support to the national sanity system, to school and to universities, as they promised all along the government of their  {\em political enemies}. In other words: they simply understand how simpler is to be part of the minority of the Parliament. Being in the majority means to govern, and govern often means to take unpopular choices. These politicians are voted, of course, by a group of electors which, at the time in which the elections occur, are their (strong) supporters. But after some time, they could lose their trust in those they voted for. And, maybe, they can completely change their opinion on the governing coalition. This is exactly what we want to model in this paper: two groups of people, a {\em political coalition} $\C$, possibly made by different parties,  and their {\em supporters} $\Sc$. At $t=0$ $\Sc$ fully trusts in $\C$. Each supporter believes that the leader of $\C$ is the perfect person to {\em rule the Country}. The supporters are sure that $\C$ will be coherent with its electoral promises, that will be efficient, and will also be very competent. We all know, unfortunately, that this is not the case in real life. In this paper we are interested in describing how this feeling changes, and what are the consequences of these changes.

The technical way in which our models will be constructed is following a {\em quantum-like} approach. Since some decades an always growing group of people started to be interested in the possibility of using quantum ideas in the analysis of macroscopic systems. This is relevant both for the {\em physical implications} of a similar approach, and for its natural mathematical framework, which can be sometimes easily and naturally connected to the system under analysis. Just to cite few applications of this quantum-like approach, this is often adopted in Decision Making to describe order effects and conjunction fallacy, in Finance to describe (simplified) stock markets, in social sciences to describe some kind of phase transition occurring in large systems (like revolutions, or some massive use of a specific social network, or the appearance of some viral news). Many other applications have also been considered along the years. We refer to \cite{abbott}-\cite{QLH} for a list of monographs where quantum-like ideas have been considered and applied to many different problems. In view of our concrete application here, it is useful to cite other quantum-like approaches to politics proposed along the years, but with different focus with respect to the one considered here. We refer to \cite{bag2015SIAP}-\cite{andro} for a series of papers considering, e.g., the possibility of using quantum tools to secure anonymous voting, or to the role of a decision making procedure when forming political coalitions, or yet the description of people changing coalition with time. We see that quantum-like ideas are widely used nowadays, mainly because they open the way to different and apparently exotic descriptions of some interesting systems, whose {\em standard description} is not fully satisfactory.

The paper is organized as follows:

In the next section we state the problem and we propose the strategy we will adopt in the rest of the paper. In particular we focus on the role of non self-adjoint (or, equivalently, non Hermitian) Hamiltonians in our quantum-like approach.

In Section \ref{sect3} we propose different such Hamiltonians and we analyze the dynamics they produce, in the context of our full system  $\Sc\cup\C$. More specifically, in Section \ref{sect3a} we introduce an unbalanced Hamiltonian\footnote{We briefly recall that a balanced Hamiltonian can be written as a sum of contributions, each one with the same number of raising and lowering operators. We refer to the Appendix for a more detailed definition.} and we analyze the {\em satisfaction} of the supporters $\Sc$ with time. This same quantity is analyzed in the other models proposed in Sections \ref{sect3b} and \ref{sect3c}. The difference is in the form of the Hamiltonian we consider, balanced or not, and describing or not rational voters. In Section \ref{sectconfronto} we briefly discuss some relations between our present approach and other possibilities proposed along the years in the literature. Section \ref{sectconcl} contains our conclusions and plans for future. To keep the paper self-contained, and in view of its relevance for this paper, we have included some results on the so-called $\gamma$-dynamics in the Appendix.

  \section{Preliminaries}\label{sect2}
  
The system $\Sigma$ we are interested in here is the union of two groups of people, those belonging to a certain {\em coalition} $\C$, and their supporters $\Sc$: $\Sigma=\C\cup\Sc$. At $t=0$ the political elections have just occurred, and $\C$ has been elected to give rise to the government $\G$ of the Country. The only two variables  we consider in our analysis are $\Cf$ and $\Ff$, describing respectively the {\em overall efficiency} of $\G$ as it is perceived by $\Sc$, and the related satisfaction\footnote{$\Ff$ stands for {\em faith} or for {\em (positive) feedback}.} of $\Sc$ as a consequence of this perception. These two features are {\em measured} by two (Hermitian and positive) operators $\ch$ and $\fh$, $\ch=\ch^\dagger$ and $\fh=\fh^\dagger$, by taking their mean values on some suitable vector in a way which will be discussed later, and which has been used in the past years by many researchers, see  \cite{bagbook,bagbook2,FFF} and references therein. For instance, if during the time evolution of $\Sigma$ the mean value of $\ch$ assumes its maximum value, then people in $\Sc$ have a good perception of what $\G$ is doing. On the other hand, if this mean value is zero, which is the minimum allowed value, then the supporters perception is that $\G$ is not efficient (or coherent) at all. At time $t=0$ we assume that $[\ch,\fh]=0$. This is useful to describe the fact that, at the time of the elections, the two quantities $\Cf$ and $\Ff$ are independent\footnote{In the language of Decision Making, $\Cf$ and $\Ff$ should be called {\em compatible}.}: people in $\Sc$ would support coalition $\C$ in any case! Even because (possibly) they have not yet a proof  of the quality and of its efficiency. Or, as it often happens, because they have {\em no memory} of what $\C$ did in the recent past.

Of course, after some days from their election, $\C$ should prove that they are truly efficient, competent, and that their choices are coherent with their electoral promises. Depending on the fact that this is true or not, $\Ff$ can change or not: the mean value of $\fh$ should stay constant (and high) if $\Sc$ are satisfied with what $\G$ is doing. Otherwise this mean value should decrease. In this perspective, $\ch$ and $\fh$ cease to be independent, so that we would expect that, calling $\ch(t)$ and $\fh(t)$ their time evolution, $[\ch(t),\fh(t)]\neq0$ for $t>0$ or, at least, for some $t>0$. However, it is easy to see that this is not possible if we use a {\em standard view} to the time evolution of the observables. In this view the time evolution is driven by an Hermitian  Hamiltonian $H_0$. If $H_0$ does not depend on time explicitly, it is well known that all observables $X\in\B(\Hil)$, the set of bounded operators on the Hilbert space $\Hil$, satisfies the Heisenberg equation of motion
\be
\dot X(t)=i[H_0,X(t)], \qquad \mbox{ so that }\qquad  X(t)=e^{iH_0t}Xe^{-iH_0t},
\label{21}\en
\cite{bagbook,mess,mer}. Hence, it is clear that, given any two commuting operators $X,Y\in\B(\Hil)$, $[X,Y]=0$, then
$$
[X(t),Y(t)]=e^{iH_0t}[X,Y]e^{iH_0t}=0,
$$
$\forall t\geq0$: two compatible observables, stay compatible at all time. This suggests that self-adjoint Hamiltonians are not the right class of operators to describe time evolution in concrete situations in which we expect that two observables which are naturally compatible at $t=0$, will evolve into not  compatible observables later on, at least for some $t>0$. There is indeed a solution for this problem, and consists in replacing a self-adjoint Hamiltonian $H_0$ with some other operator, which we still call {\em Hamiltonian}, $H$, which is {\bf not} self-adjoint: $H\neq H^\dagger$. There is a huge literature on Hamiltonians of this kind, since it appeared clear that some of these Hamiltonians can be efficiently used in the description of open or gain-and-loss systems. We refer to \cite{benbook}-\cite{mois}  for some monographs and edited volumes on this topic. Thousands of scientific papers can be easily  found in the references of these books, or just looking on the web. In this paper we are only interested in some specific aspects of these Hamiltonians, and in particular in the dynamics which is generated by an $H\neq H^\dagger$, and on the existence of conserved quantities. These aspects are discussed in many papers, \cite{sgh}-\cite{lisok}, but we refer here to \cite{bag2022,bag2025,baglia}, where the interest is exactly on those aspects which are relevant for us here. We also refer to the Appendix, where we list some essential results used all along the paper. 

If the Hamiltonian of a certain system is not self-adjoint, $H\neq H^\dagger$, and does not depend explicitly on time, the Heisenberg time evolution of each operator $X$ of the system is $\gamma^t(X)=e^{iH^\dagger t}Xe^{-iHt}$, see the Appendix. It is clear that $\gamma^t(XY)\neq\gamma^t(X)\gamma^t(Y)$, in general, and therefore $[\gamma^t(X),\gamma^t(Y)]$ can easily be non zero (for some or all $t>0$) even if $[X,Y]=0$: two compatible observables may  become incompatible while they evolve in time. This is exactly what we would like to have: even if $[\ch,\fh]=0$, we expect that the time evolution of this commutator is non zero, at least for some time $t>0$. This is why the role of the dynamics $\gamma^t$, the so-called $\gamma$-dynamics, is so relevant for us here.

\vspace{2mm}

{\bf Remarks:--} (1) Of course it is easy to extend what we have discussed so far to the case in which, for some reason, the two observables $X$ and $Y$ ($\hat C$ and $\hat F$ in our case), are not independent already at $t=0$. It is sufficient to assume that $[\gamma^0(X),\gamma^0(Y)]=[X,Y]\neq0$, and check what this commutator becomes for $t\geq0$. It might happen, in principle, that for some $t_0>0$, $[\gamma^{t_0}(X),\gamma^{t_0}(Y)]=0$: two originally incompatible observables become, at some specific time, compatible! Also this situation cannot be easily described in terms of self-adjoint Hamiltonians.

(2) It is important to stress that the possibility of using non self-adjoint Hamiltonians to describe changing nature observables is only possible using the Heisenberg representation. In the Schr\"odinger representation, when the wave functions evolve in time, while the observables do not, it is clear that we cannot describe a similar behaviour: each pair of, say, compatible observables, remain compatible! This remark suggests once more the relevance of $\gamma^t$ in applications.

\vspace{2mm}

Following what discussed in the Appendix, we claim  that, given an observable $X\in\B(\Hil)$ of $\Sigma$, the most relevant quantity to compute is $\hat x(t)=\pin{\hat\Psi(t)}{X\hat\Psi(t)}$, where $\hat\Psi(t)=\frac{\Psi(t)}{\norm{\Psi(t)}}$ is the normalized wave function for $\Sigma$, and $\Psi(t)$ obeys the Schr\"odinger equation $i\dot\Psi(t)=H\Psi(t)$ or, equivalently, $\Psi(t)=e^{-iHt}\Psi(0)$, with $\Psi(0)$ describing $\Sigma$ at $t=0$. In other words, what are really relevant for us here are the following functions:
\be
\hat c(t)=\pin{\hat\Psi(t)}{\hat C\hat\Psi(t)}, \qquad\qquad \hat f(t)=\pin{\hat\Psi(t)}{\hat F\hat\Psi(t)},
\label{22}\en
see (\ref{a4}). These functions are bounded for all $t\geq0$. Indeed a simple use of the Cauchy-Schwarz inequality produces the following bounds: $|\hat c(t)|\leq\|\ch\|$ and $|\hat f(t)|\leq\|\fh\|$, $\forall t\geq0$.

Other quantities which are useful to compute, also in view of their physical interpretation, are the variance of $\ch$ and $\fh$. These variances, how we will discuss later, will be related to a sort of uncertainty of $\Sc$ in their feeling for $\C$. Due to the relevance that non Hermitian operators have for us, we first observe that the standard formula $(\Delta X)^2=\langle X^2\rangle-\langle X\rangle^2$, does not hold for a generic $X\neq X^\dagger$. In this case it should be replaced by
\be
(\Delta X)^2=\|(X-\langle X\rangle)\varphi\|^2=\langle X^\dagger X\rangle_\varphi-\langle X^\dagger \rangle_\varphi\langle X\rangle_\varphi,
\label{23}\en
see e.g. \cite{bagunc}, where we use the notation $\langle X\rangle_\varphi=\pin{\varphi}{X\varphi}$, for instance. It is clear that we go back to the previous formula if $X=X^\dagger$. Notice that in $\Delta X$ we are not writing explicitly the dependence on $\varphi$. We will often make this choice, except when this creates confusion.

\vspace{2mm}

{\bf Remark:--} If we restrict to $X=\hat C$ or to $X=\hat F$ there would be no need to use (\ref{23}) rather than the usual one, since it is possible to check that $(\gamma^t(\ch))^\dagger=\gamma^t(\ch^\dagger)=\gamma^t(\ch)$ and $(\gamma^t(\fh))^\dagger=\gamma^t(\fh^\dagger)=\gamma^t(\fh)$ for all $t\geq0$: the $\gamma$-dynamics does not destroy the self-adjointness of $\ch$ and $\fh$. Of course, (\ref{23}) can still be used also for self-adjoint operators, giving back the standard formula. As already mentioned, in our context, where non Hermitian operators play a significant role, (\ref{23}) is the natural formula to adopt for variances. 

\vspace{2mm}

From  (\ref{23}) we get
$$
(\Delta \gamma^t(X))^2=\|\gamma^t(X)\varphi\|^2-|\langle\gamma^t(X)\rangle_\varphi|^2,
$$
which,  with simple manipulations, can be rewritten as
$$
(\Delta \gamma^t(X))^2=\|e^{iH^\dagger t}X\Psi(t)\|^2-|\langle X\rangle_{\Psi(t)}|^2,
$$
where $\Psi(t)=e^{-iHt}\varphi$. In coherence with our choice to use normalized vectors, see (\ref{22}), we define
\be
(\Delta_n \gamma^t(X))^2=\left|\|e^{iH^\dagger t}X\hat\Psi(t)\|^2-|\langle X\rangle_{\hat\Psi(t)}|^2\right|,
\label{24}\en
for generic $X$. The role of the absolute value in this formula is that, first of all, replacing $\Psi(t)$ with $\hat \Psi(t)$, $(\Delta \gamma^t(X))^2$ would be replaced by a quantity (if we don't put the absolute value) which is not necessarily positive, in contrast with what we expect a variance should be.

\vspace{2mm}

{\bf Remark:--} It may be useful to stress that, for self-adjoint Hamiltonians, (\ref{24}) is exactly the same as the standard variance, \cite{mer,mess}. In particular, there is no need of taking the absolute value in $(\Delta_n \gamma^t(X))^2$ to ensure that this quantity is positive. So, in a sense, (\ref{24}) is to be understood as a {\em natural proposal}. To make this proposal more robust, we plan to consider further this definition in more applications in the future.

\vspace{2mm}

Next we observe that, since
  $$
  \left\|e^{\pm iHt}\right\|\leq e^{|t|\|H\|}, \qquad\mbox{and}\qquad  \left\|e^{\pm iH^\dagger t}\right\|\leq e^{|t|\|H\|},
  $$
  then 
  \be
   (\Delta_n \gamma^t(X))\leq\|X\|\left(e^{2|t|\|H\|}+1\right)^{1/2},
   \label{25}\en
   which gives an useful bound for the variance of the generic operator $X$.
  In particular this formula shows that, for $t=0$, $(\Delta_n \gamma^t(X))\leq\sqrt{2}\,\|X\|$.

  To analyze the dynamics of $\Sigma$, and in particular what happens to $\Ff$, the quantities we are mostly interested here are $\hat f(t)$ and the related variance, $(\Delta_n \gamma^t(\fh))$. The natural interpretation is the following:
  
  \begin{itemize}
  	\item  If for some $t$ we find $\hat f(t)\simeq1$ and $(\Delta_n \gamma^t(\fh))\simeq0$,
  	we conclude that $\Sc$, at that time $t$, is still very happy with $\G$, without any particular perplexity. Stated differently: $\Sc$'s feedback is very high, with almost zero uncertainty.

  	\item  If for some $t$ we rather have $\hat f(t)\simeq1$ and a "large" (according to the bound in (\ref{25})) $(\Delta_n \gamma^t(\fh))$,
  	then $\Sc$ is still  happy with $\G$, but with some (or many) perplexities.
  	
  	\item  If for some $t$ we rather have $\hat f(t)\simeq0$ and a "large" (again, according to (\ref{25})) $(\Delta_n \gamma^t(\fh))$,
  	then $\Sc$ is not  happy with $\G$ at all, but it is still expecting a possible improvement in $\G$'s action.
  	
  	\item  lastly, if for some $t$ we have $\hat f(t)\simeq0$ and a $(\Delta_n \gamma^t(\fh))\simeq 0$,
  	then $\Sc$ is not  happy with $\G$ at all, and it does not expect any improvement in $\G$'s action. Of course, this is the worst situation for $\C$, since its original supporters are no longer supporting $\G$.

  \end{itemize}

What we have discussed so far is rather general. In the next section we will propose different models, with balanced or unbalanced Hamiltonians, see \cite{baglia}, and we will discuss their dynamics and check the differences between these different proposals.

\section{Models}\label{sect3}

In this section we propose three different models, all driven by non self-adjoint Hamiltonians described in terms of fermionic ladder operators. In the first two sections the Hamiltonians model a {\em rational} behavior: the electors' feedback moves in line with the perceived efficiency, while in the third Hamiltonian this is not the case: the Hamiltonian $H_3$ will contain rational and {\em irrational} terms, and we will discuss the competition between the two.

\subsection{First Hamiltonian}\label{sect3a}

Here and in the following the main ingredients of our models are the operators $c$ and $f$ which, together with their adjoints $c^\dagger$ and $f^\dagger$, satisfy the following canonical anti-commutation relations (CAR):
\be
\{c,c^\dagger\}=\{f,f^\dagger\}=\1, \qquad \{c^\sharp,f^\sharp\}=0, \qquad c^2=f^2=0.
\label{31}\en
Here $\{a,b\}=ab+ba$ is the anti-commutator between the operators $a$ and $b$, and $a^\sharp$ stands for $a$ or $a^\dagger$. The operators $\ch$ and $\fh$ which, in the previous section, were rather general are now fixed to be those which are known in the literature as the {\em number operators} for the two fermionic modes: $\ch=c^\dagger c$ and $\fh=f^\dagger f$. These two latter operators are self-adjoint. If we now introduce the vacuum of $c$ and $f$, i.e. the non zero vector in $\Hil=\mathbb{C}^4$ satisfying $c\,\varphi_{0,0}=f\,\varphi_{0,0}=0$, we can further construct an orthonormal (o.n.) basis for $\Hil$ in a standard way:
\be
\varphi_{1,0}=f^\dagger\varphi_{0,0}, \qquad \varphi_{0,1}=c^\dagger\varphi_{0,0}, \qquad \varphi_{1,1}=f^\dagger c^\dagger\varphi_{0,0}.
\label{32}\en 
Notice that, because of the CAR, $\varphi_{1,1}=-c^\dagger f^\dagger\varphi_{0,0}$: the order of the operators can be relevant. Notice that, see \cite{bagbook,bagbook2,FFF}, these vectors have a simple (and natural) interpretation, due to the fact that $\fh\varphi_{n,k}=n\varphi_{n,k}$ and $\ch\varphi_{n,k}=k\varphi_{n,k}$: $\varphi_{1,1}$ describes the fact that $\Sc$ is perceiving an high efficiency of $\G$, and returns a very positive feedback. The vector $\varphi_{0,0}$ describes the opposite situation: $\Sc$ believes that $\G$ is not working well, and the feedback is very negative. And so on.

The Hamiltonian we consider here is the following:
\be
H_1=\lambda_1 c^\dagger f^\dagger +\lambda_2 fc,
\label{33}\en
where $\lambda_1$ and $\lambda_2$ are strictly positive constants. Notice that if $\lambda_1=\lambda_2$, then $H_1=H_1^\dagger$. Otherwise this is not true. The two terms in $H_1$ can be interpreted as follows: $c^\dagger f^\dagger$ describes the fact that, if the perceived efficiency of $\G$ increases (because of $c^\dagger$), the approval of $\Sc$ for $\C$ increases as well (thanks to $f^\dagger$). The term $fc$ describes the opposite effect: if the perceived efficiency of $\G$ decreases, the approval of $\Sc$ for $\C$ also decreases. This is what we have called {\em a rational} Hamiltonian. Indeed, it is exactly how one  expect some rational voter would behave.

In order to compute $\hat f(t)$ and $(\Delta_n \gamma^t(\fh))$ the first thing to do is to compute $\Psi(t)=e^{-iH_1t}\Psi(0)$, and, in particular, $e^{-iH_1t}$. As for $\Psi(0)$, in view of what we have discussed previously, it is natural to assume that $\Psi(0)=\varphi_{11}$, since this is the vector which describes the maximum perceived efficiency of $\G$ and the maximum approval of $\Sc$ for $\G$. 

The computation of $e^{-iH_1t}$ is a nice exercise which is based on the following results: calling $A:=2\hat F\hat C-\hat C - \hat F+\1$, it is possible to prove, using induction on $n$, that
\be
H_1^{2n}=(\lambda_1\lambda_2)^nA, \qquad \mbox{and}\qquad  H_1^{2n+1}=(\lambda_1\lambda_2)^n H_1,
\label{35}\en
$\forall n\geq1$. Using these equalities  we find
\be
e^{-itH_1}=\1-\frac{i}{\sqrt{\lambda_1\lambda_2}}\sin(t\sqrt{\lambda_1\lambda_2})\,H_1+\left(\cos(t\sqrt{\lambda_1\lambda_2})-1\right)A.
\label{36}\en
Taking the adjoint of (\ref{36}) we deduce that
\be
e^{itH_1^\dagger}=\1+\frac{i}{\sqrt{\lambda_1\lambda_2}}\sin(t\sqrt{\lambda_1\lambda_2})\,H_1^\dagger+\left(\cos(t\sqrt{\lambda_1\lambda_2})-1\right)A.
\label{37}\en
Then the wave function $\Psi(t)=e^{-iH_1t}\varphi_{11}$ looks as
\be
\Psi(t)=i\sqrt{\frac{\lambda_2}{\lambda_1}}\sin(t\sqrt{\lambda_1\lambda_2})\,\varphi_{00}+\cos(t\sqrt{\lambda_1\lambda_2})\,\varphi_{11},
\label{38}\en
which implies that
\be
\|\Psi(t)\|^2=\frac{\lambda_2}{\lambda_1}\sin^2(t\sqrt{\lambda_1\lambda_2})+\cos^2(t\sqrt{\lambda_1\lambda_2}).
\label{39}\en
This formula shows that, if $\lambda_1=\lambda_2$, then $\|\Psi(t)\|=1$ for all $t$, in agrement with the fact that, under this condition, $H_1=H_1^\dagger$ and therefore $e^{-iH_1t}$ is unitary, and, as such, preserves the norms.

\begin{figure}[htbp]
    \centering
    
    \begin{subfigure}{0.45\textwidth}
        \centering
        \includegraphics[width=\linewidth]{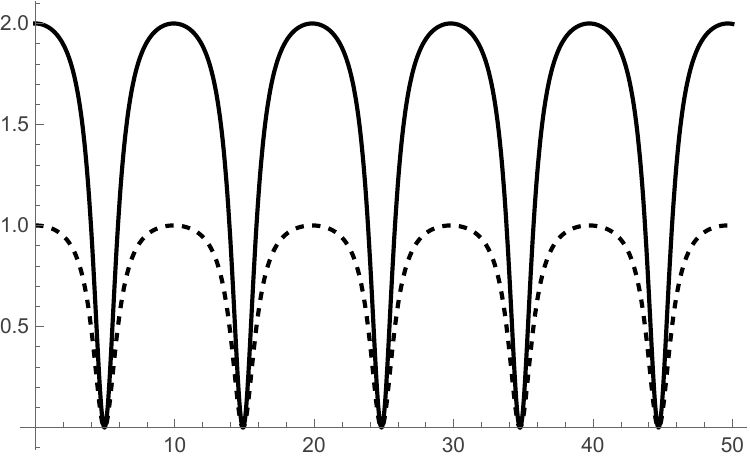}
    \end{subfigure}
    \hfill
    \begin{subfigure}{0.45\textwidth}
        \centering
        \includegraphics[width=\linewidth]{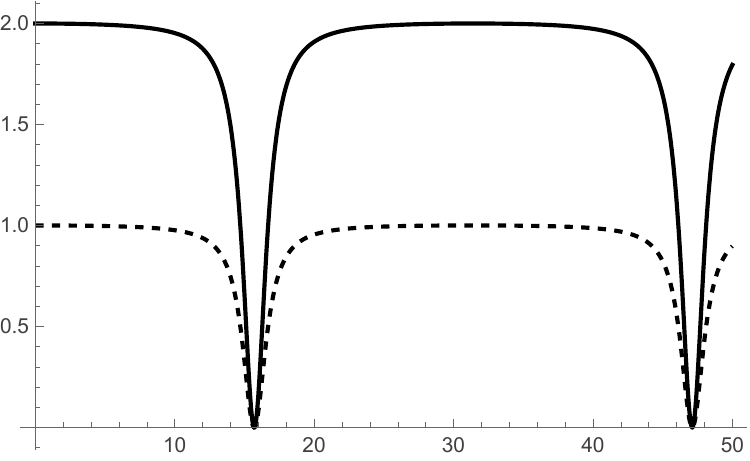}
    \end{subfigure}
    
    \vspace{0.5cm}
    
    \begin{subfigure}{0.45\textwidth}
        \centering
        \includegraphics[width=\linewidth]{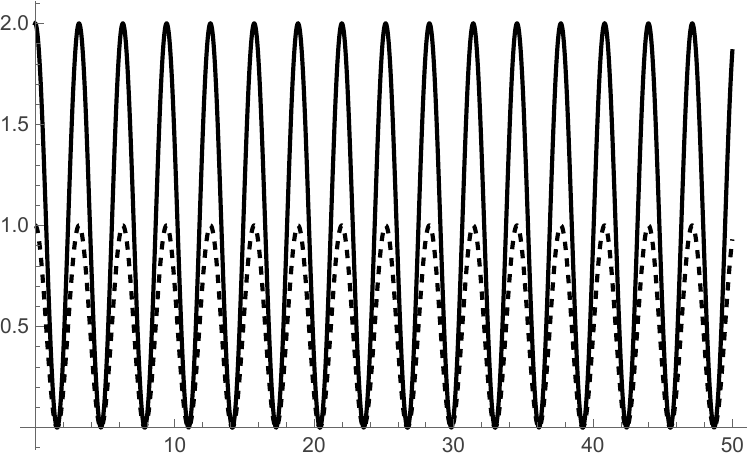}
    \end{subfigure}
    
    \caption{$\hat f(t)=\hat c(t)$ (dashed line) and their sum (thick line), for $\Psi(0)=\varphi_{11}$ and $(\lambda_1, \lambda_2)=(1,0.1)$ (up left), $(\lambda_1, \lambda_2)=(1, 0.01)$ (up right), $(\lambda_1, \lambda_2)=(1,1)$ (down) }
    \label{fig1}
\end{figure}

We have now all the ingredients to compute the quantities we are interested in, i.e. the (normalized) mean values in (\ref{22}) and the (again, normalized) variances in (\ref{24}), with $X=\ch,\fh$. Due to the specific initial condition we are considering ($\Psi(0)=\varphi_{11}$), and to the form of the Hamiltonian, is not a big surprise to find that $\fh$ and $\ch$ behave in the same way. Indeed we find that
\be
\hat f(t)=\hat c(t)=\frac{\cos^2(t\sqrt{\lambda_1\lambda_2})}{\frac{\lambda_2}{\lambda_1}\sin^2(t\sqrt{\lambda_1\lambda_2})+\cos^2(t\sqrt{\lambda_1\lambda_2})}.
\label{310}\en
As we expected, since $H_1$ is not {\em balanced}, \cite{baglia}, $\hat f(t)+ \hat c(t)$ is not a conserved quantity.\\ Figure \ref{fig1} shows the plots of $\hat f(t)$, $\hat c(t)$ and their sum, for different values of $\lambda_1$ and $\lambda_2$. In particular we plot these functions for $\lambda_1$ larger that $\lambda_2$  (up left), for $\lambda_1$ much larger that $\lambda_2$ (up right), and  for $\lambda_1=\lambda_2$ (down).
\\
This is what we expected, indeed, if the evolution is driven by an Hamiltonian in which the weight of the creation term ($c^\dagger f^\dagger$) is larger than the other when starting from $\Psi(0)=\varphi_{11}$. In fact, this initial condition implies that, at $t=0$, electors' perceived efficiency and satisfaction are maximum,  so that $\hat f(0)=\hat c(0)=1$. Then, since $\lambda_1\geq \lambda_2$, it is natural to expect that  $\hat f(t)$ and $\hat c(t)$ will {\em stay close} to $1$ for (almost) all time\footnote{They cannot exceed 1, due to the fermionic nature of $\ch$ and $\fh$.}. They move away from the value 1 only in some time intervals, whose length become shorter and shorter as soon as $\lambda_2$ decreases when compared to $\lambda_1$. Of course, we expect the opposite happens, i.e. $\hat f(t)$ and $\hat c(t)$ stay close to $0$,  if we decrease the value of $\lambda_1$ when compared to $\lambda_2$. The case $\lambda_1=\lambda_2$ produces exactly the plot we expect to find: pure oscillations for $\hat f(t)$ and $\hat c(t)$. This is in agreement with a well known no-go result, \cite{bagbook}: in presence of a self-adjoint Hamiltonian, acting on a finite-dimensional Hilbert space, we can only deduce a simple oscillatory motion.

It is easy to observe that choosing as initial state $\Psi(0)=\varphi_{10}$ or $\Psi(0)=\varphi_{01}$, $\hat f(t)+ \hat c(t)$ turns out to be constant. This happens because these two states will not evolve in time, as we can easily deduce using (\ref{36}). This result can be understood as follows: suppose, for instance, that at $t=0$ the electors are in a state $\Psi(0)=\varphi_{10}$. This means that they perceive  minimum efficiency of $\G$, but nevertheless they return a positive feedback\footnote{Maybe because they think $\G$ could do worst, or maybe because $\Sc$ has other interests that $\G$ is still not {\em touching}.}. We see that there is a sort of contradiction between these initial conditions and the "rational" nature of $H_1$. This contradiction produces, as its effect, a constant value of $\hat f(t)+ \hat c(t)$ in this case. A similar conclusion can be deduced if $\Psi(0)=\varphi_{01}$.

The variances are:

\be
(\Delta_n \gamma^t(X))=\frac{1}{2}|\sin(2t\sqrt{\lambda_1\lambda_2})|\sqrt{\frac{\sin^2(t\sqrt{\lambda_1\lambda_2})+\cos^2(t\sqrt{\lambda_1\lambda_2})\frac{\lambda_1^2+\lambda_2^2-\lambda_1\lambda_2}{\lambda_1\lambda_2}}{(\frac{\lambda_2}{\lambda_1}\sin^2(t\sqrt{\lambda_1\lambda_2})+\cos^2(t\sqrt{\lambda_1\lambda_2}))^2}},
\en
where $X=\hat F, \hat C$.\\
These formulas show in particular what follows:

\begin{enumerate}
	\item if $t=\frac{\pi}{2\sqrt{\lambda_1\lambda_2}}$, then $\hat f(t)=0$, and $(\Delta_n \gamma^t(\fh))=0$ as well. As we have discussed before, this implies that there is a minimum of the feedback of $\Sc$, with minimum doubt. This is the worst case for $\G$, of course.
	
	\item if $t=\frac{\pi}{\sqrt{\lambda_1\lambda_2}}$, then $\hat f(t)=1$, and $(\Delta_n \gamma^t(\fh))=0$. In this case $\Sc$ has a very good opinion on $\G$'s action, again with no doubt.
	
	\item if $t=\frac{3\pi}{2\sqrt{\lambda_1\lambda_2}}$, then $\hat f(t)=(\Delta_n \gamma^t(\fh))=0$, and we go back to the first situation.
\end{enumerate}

We see that the reaction of $\Sc$ has a periodic behavior, and the period is related to the product of the $\lambda$'s in $H_1$. This is indeed what we observe in politics, where there is often some alternation between, say, left- and right-wing parties, over time. 

From (\ref{310}) we also observe that:

\begin{itemize}
	\item if $\lambda_1=\lambda_2=:\lambda$ then
	\be
	\hat f(t)=\hat c(t)=\cos^2(t\lambda),\qquad \mbox{and}\qquad (\Delta_n \gamma^t(\ch))=(\Delta_n \gamma^t(\fh))=\frac{1}{2}|\sin(2t\lambda)|
	\label{312}\en
	\item if $\lambda_2\gg\lambda_1$ then
	\be
		\hat f(t)=\hat c(t)\simeq0,\qquad \mbox{and}\qquad (\Delta_n \gamma^t(\ch))=(\Delta_n \gamma^t(\fh))\simeq0.
		\label{314}\en
	This is the worst result for $\G$: the supporters are not supporting $\C$ anymore, and they are very unsatisfied, with very few doubts.
	\item if $\lambda_2\ll\lambda_1$ then
	\be
	\hat f(t)=\hat c(t)\simeq1,\qquad \mbox{and}\qquad (\Delta_n \gamma^t(\ch))=(\Delta_n \gamma^t(\fh))\simeq \sqrt{\frac{\lambda_1}{\lambda_2}}|\sin(2t\sqrt{\lambda_1\lambda_2})|.
	\label{315}\en
	We see that, even if $\hat f(t)\simeq 1$, the variance of $\fh$ can change with time: $\Sc$ are still satisfied by $\C$, but they start to wonder if something is going wrong. However, as Figure \ref{fig1} shows, this is the best situation for $\C$ since the time intervals in which $\hat f(t)$ is significantly different from 1 are really short.

\end{itemize}

According to these results the ratio $\nu=\frac{\lambda_1}{\lambda_2}$ can be interpreted as an {\em index of success} for $\G$: the larger $\nu$, the better the situation for $\G$: their supporters may have some doubts, but they are still very happy with what $\G$ is doing. On the other hand, if $\nu\simeq0$, then $\Sc$ is not really appreciating $\G$ anymore. Going back to $H_1$ in (\ref{33}), this is exactly what we could expect, since when, say, $\nu\simeq0$, then the term $fc$ is {\em more efficient} than $c^\dagger f^\dagger$, which implies that $\Sc$ is changing drastically its original idea on $\C$ and $\G$.

We see that, apparently, there is no conserved quantity in general, and in particular for our choice of $\Psi(0)$, at least when we focus on $\hat f(t)$ and $\hat c(t)$. This is in agreement with the fact that $H_1$ is not balanced, see \cite{baglia}. We will go back on this aspect in the next section.

We end this section by showing that the computation of $[\gamma^t(\fh),\gamma^t(\ch)]$ produces an apparently unexpected result. In fact, as we have discussed in Section \ref{sect2}, it is quite plausible that, given two operators $X$ and $Y$ such that $[X,Y]=0$, then $[\gamma^t(X),\gamma^t(Y)]\neq0$. This is not what happens here, at least for $\ch$ and $\fh$. Indeed some straightforward computations produce

\be
\gamma^t(\fh)=X[\fh,\ch]+\frac{i}{2}\sqrt{\frac{\lambda_1}{\lambda_2}}\sin(2t\sqrt{\lambda_1\lambda_2})\left(fc-c^\dagger f^\dagger\right),
\label{316}\en
and
\be
\gamma^t(\ch)=X[\ch,\fh]+\frac{i}{2}\sqrt{\frac{\lambda_1}{\lambda_2}}\sin(2t\sqrt{\lambda_1\lambda_2})\left(fc-c^\dagger f^\dagger\right),
\label{317}\en
where we have introduced the following operator-valued function:
$$
X[\fh,\ch]=\fh\left(\1-\frac{\lambda_1}{\lambda_2}\sin^2(t\sqrt{\lambda_1\lambda_2})\right)+\frac{\lambda_1}{\lambda_2}\sin^2(t\sqrt{\lambda_1\lambda_2})(\1-\ch)+
$$
\be
+\fh\ch\sin^2(t\sqrt{\lambda_1\lambda_2})\left(\frac{\lambda_1}{\lambda_2}-1\right).
\label{318}\en
Then it is clear that
\be
[\gamma^t(\fh),\gamma^t(\ch)]=0
\label{319}\en
for all $t\geq0$. Then a natural question arises: is there any hidden reason for this to happen? Indeed the reason exists, and it is really not so hidden. The point is, we believe, 
that the two observables $\hat F$ and $\hat C$ remain compatible for all $t\geq0$ because the Hamiltonian $H_1$ contains terms that increase and decrease the two quantum numbers of the observables $\hat F$ and $\hat C$ {\bf together}: stated differently, $\hat F$ and $\hat C$ are two copies of essentially the same dynamical quantity, which (moreover) satisfy the same initial condition if $\Psi(0)=\varphi_{11}$. Hence there is no real reason why, with an Hamiltonian like $H_1$, they should become incompatible after some time. We will show in the next section how to modify the model in order to get compatible quantities which evolve into incompatible operators.

\subsection{Second Hamiltonian}\label{sect3b}
We consider here the same mathematical framework, but we rather adopt a balanced Hamiltonian, which we again construct  modeling a rational behavior. For doing that, we need to replace the operators $f,f^\dagger$ and $\fh$ with $r,r^\dagger$ and $\rh=r^\dagger r$ whose interpretation are connected but opposite: while $f^\dagger$ describes an increasing feedback of $\Sc$ for $\C$, $r^\dagger$ describes an increasing {\em lack of approval} of $\Sc$ for $\C$ (or, stated differently, a decreasing positive feedback).  Let us now briefly comment on the meaning of the eigenstates constructed (in analogy with the previous case) for these new operators. Differently from the previous section, the first quantum number of the vector $\varphi_{R,C}$ represents the eigenvalues of $\hat R$ and, accordingly to its interpretation, the eigenvalue "$0$" represents when the electors are not un-satisfied (i.e., the electors are satisfied) and "$1$" represents when the electors are fully un-satisfied. In other words, in terms of interpretation, $r^\dagger$ will be here what $f$ was in $H_1$. And $r$ will play the same role as $f^\dagger$ before, and the initial state is now $\Psi(0)=\varphi_{0,1}$, since this describes minimum un-satisfaction  and maximum perceived efficiency, in analogy to what we have assumed in Section \ref{sect3a}.
The expression of $H_2$, which describes the same phenomena as $H_1$ is:
\be
H_2=\lambda_1 c^\dagger r +\lambda_2 r^\dagger c,
\label{320}\en
However, there is still a big (mathematical) difference between $H_1$ and $H_2$, since this latter is balanced, while $H_1$ is not. This implies that, see \cite{baglia}, some specific mean values are indeed constant in time. More specifically, we expect to find that
\be
\hat c(t)+\hat r(t)=\mbox{const},
\label{321}\en
where, as in (\ref{22}), we put  $\hat r(t)=\pin{\hat\Psi(t)}{\hat R\hat\Psi(t)}$.\\

Introducing the operator $B:=\hat{R}+\hat{C}-2(\hat{R}\hat{C})$ and following the same steps as for $H_1$, we find that
\be
H_2^{2n}=(\lambda_1\lambda_2)^nB \qquad \mbox{and}\qquad H_2^{2n+1}=(\lambda_1\lambda_2)^nH_2,\en
$\forall n\geq1$. Hence
\be
e^{-itH_2}=\1-\frac{i}{\sqrt{\lambda_1\lambda_2}}\sin(t\sqrt{\lambda_1\lambda_2})\,H_2+\left(\cos(t\sqrt{\lambda_1\lambda_2})-1\right)B,\en
\be
e^{itH^{\dagger}_2}=\1+\frac{i}{\sqrt{\lambda_1\lambda_2}}\sin(t\sqrt{\lambda_1\lambda_2})\,H_2^{\dagger}+\left(\cos(t\sqrt{\lambda_1\lambda_2})-1\right)B.\en
The wave function
$\Psi(t)=e^{-iH_2t}\varphi_{01}$ produces
\be
\Psi(t)=-i\sqrt{\frac{\lambda_2}{\lambda_1}}\sin(t\sqrt{\lambda_1\lambda_2})\,\varphi_{10}+\cos(t\sqrt{\lambda_1\lambda_2})\,\varphi_{01},\en
whose square norm is:
\be
\|\Psi(t)\|^2=\frac{\lambda_2}{\lambda_1}\sin^2(t\sqrt{\lambda_1\lambda_2})+\cos^2(t\sqrt{\lambda_1\lambda_2}).\en
In this case:
\be \hat r(t)=\frac{\frac{\lambda_2}{\lambda_1}\sin^2(t\sqrt{\lambda_1\lambda_2})}{\frac{\lambda_2}{\lambda_1}\sin^2(t\sqrt{\lambda_1\lambda_2})+\cos^2(t\sqrt{\lambda_1\lambda_2})} \quad \mbox{and}\quad \hat c(t)=\frac{\cos^2(t\sqrt{\lambda_1\lambda_2})}{\frac{\lambda_2}{\lambda_1}\sin^2(t\sqrt{\lambda_1\lambda_2})+\cos^2(t\sqrt{\lambda_1\lambda_2})}.\en\\
As we expected, their sum is conserved, indeed: $\hat r(t)+\hat c(t)=1$. This is a consequence of $H_2$ being balanced.

\begin{figure}[h!]
    \centering
    
    \begin{subfigure}{0.45\textwidth}
        \centering
        \includegraphics[width=\linewidth]{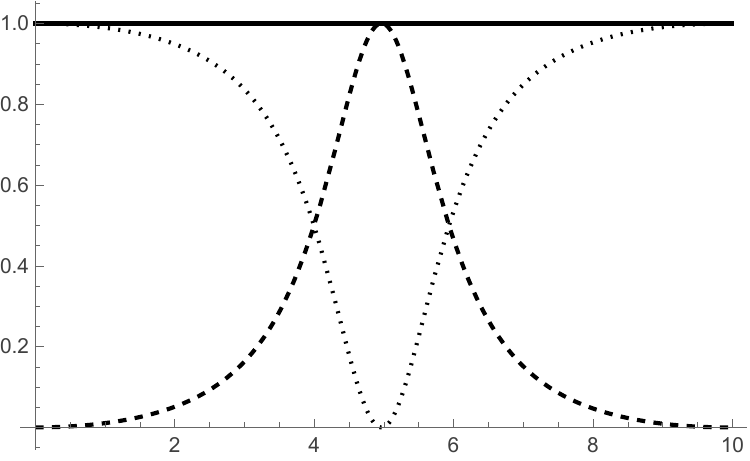}
    \end{subfigure}
    \hfill
    \begin{subfigure}{0.45\textwidth}
        \centering
        \includegraphics[width=\linewidth]{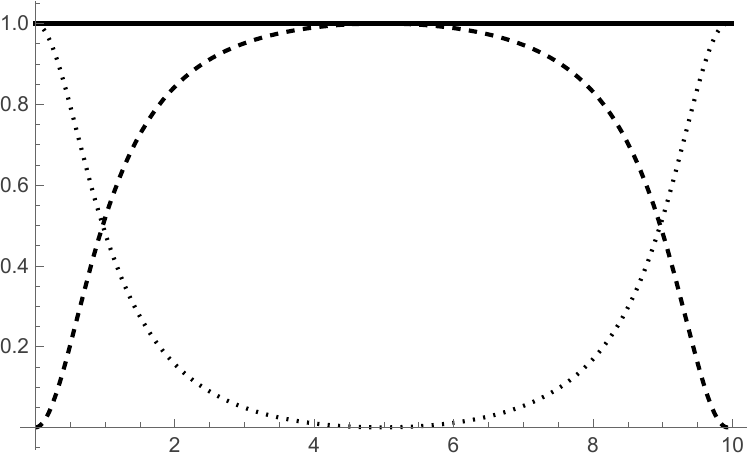}
    \end{subfigure}
    
    \vspace{0.5cm}
    
    \begin{subfigure}{0.45\textwidth}
        \centering
        \includegraphics[width=\linewidth]{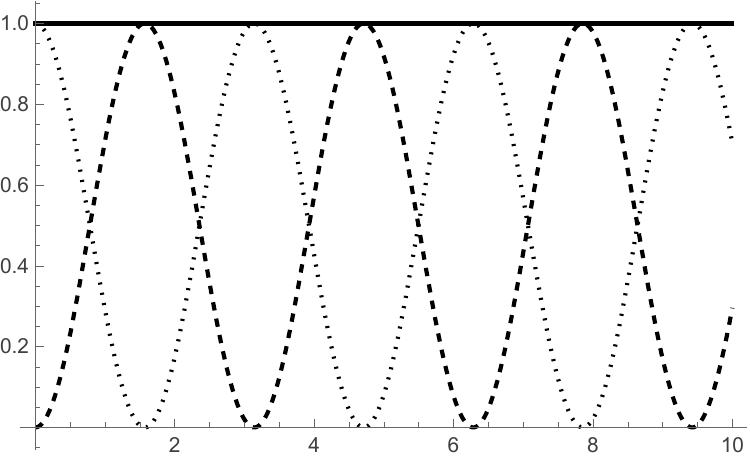}
    \end{subfigure}
    
    \caption{$\hat r(t)$ (dashed line), $\hat c(t)$ (dotted line) and their sum (thick line), for $\Psi(0)=\varphi_{01}$ and $(\lambda_1, \lambda_2)=(1, 0.1)$ (up left), $(\lambda_1, \lambda_2)=(0.1,1)$ (up right) and $(\lambda_1, \lambda_2)=(1, 1)$ (down) }
\label{fig2}\end{figure}

Figure \ref{fig2} shows the plot of $\hat r(t)$ and $\hat c(t)$ and of their sum for different values of $\lambda_1$ and $\lambda_2$. The main difference with respect to what we have found for $H_1$ is that a conserved quantity does exist, $\hat r(t)+\hat c(t)$. Still, the main other features of $\hat r(t)$ and $\hat c(t)$ are similar to the ones observed in Figure \ref{fig1}, even if, in this case, $\hat c(t)$ and $\hat r(t)$ are different functions. In particular we observe out-of-phase oscillations when $\lambda_1=\lambda_2$, in agreement with the fact that, in this case, $H_2=H_2^\dagger$. 
\\
The variances are:
\be
(\Delta_n \gamma^t(\ch))=\frac{1}{2}\left|\sin(2t\sqrt{\lambda_1\lambda_2})\right|\sqrt{\frac{\sin^2(t\sqrt{\lambda_1\lambda_2})+\cos^2(t\sqrt{\lambda_1\lambda_2})\frac{\lambda_1^2+\lambda_2^2-\lambda_1\lambda_2}{\lambda_1\lambda_2}}{(\frac{\lambda_2}{\lambda_1}\sin^2(t\sqrt{\lambda_1\lambda_2})+\cos^2(t\sqrt{\lambda_1\lambda_2}))^2}},
\en
and
\be
{(\Delta_n \gamma^t(\hat R))=\sqrt{\left|\frac{\lambda_2}{\lambda_1}\sin^2(t\sqrt{\lambda_1\lambda_2})\left(1-\frac{\frac{\lambda_2}{\lambda_1}\sin^2(t\sqrt{\lambda_1\lambda_2})}{(\frac{\lambda_2}{\lambda_1}\sin^2(t\sqrt{\lambda_1\lambda_2})+\cos^2(t\sqrt{\lambda_1\lambda_2}))^2}\right)\right|}}.
\en 
From these formulas we deduce that:

\begin{enumerate}
	\item if $t=\frac{\pi}{2\sqrt{\lambda_1\lambda_2}}$, then $\hat r(t)=1$, $\hat c(t)=0$ and {$(\Delta_n \gamma^t(\hat R))=\sqrt{\left|\frac{\lambda_2-\lambda_1}{\lambda_1}\right|}$}, $(\Delta_n \gamma^t(\ch))=0$. In this case the unsatisfaction reaches the maximum value with a doubt that depend on the values of $\lambda_{1,2}$, while the perceived efficiency is minimum, with minimum doubt. This is the worst situation for $\G$, since its supporters are very unsatisfied. Notice that this is what happened (for the same $t$) also in Section \ref{sect3a}, with a difference on the variance: in the present situation, infact, $(\Delta_n \gamma^t(\hat R))>0$ so there is still some uncertainty on the unsatisfaction of $\Sc$.
	
	\item if $t=\frac{\pi}{\sqrt{\lambda_1\lambda_2}}$, then $\hat r(t)=0$, $\hat c(t)=1$ and $(\Delta_n \gamma^t(\hat R))=0$ and $(\Delta_n \gamma^t(\ch))=0$. In this case $\Sc$ has a very good opinion on $\G$'s action, with no doubt.
	
	\item if $t=\frac{3\pi}{2\sqrt{\lambda_1\lambda_2}}$, then $\hat r(t)=1$, $\hat c(t)=0$, {$(\Delta_n \gamma^t(\hat R))=\sqrt{\left|\frac{\lambda_2-\lambda_1}{\lambda_1}\right|}$}, $(\Delta_n \gamma^t(\ch))=0$ and we go back to the first situation.
\end{enumerate}

Also here we see that the reaction of $\Sc$ is periodic, and the period is related to the product of the $\lambda$'s in $H_2$. We can also observe that:

\begin{itemize}
	\item if $\lambda_1=\lambda_2=:\lambda$ then
	\be
	\hat r(t)=\sin^2(t\lambda), \qquad\hat c(t)=\cos^2(t\lambda),\qquad \mbox{and}\qquad (\Delta_n \gamma^t(\ch))=(\Delta_n \gamma^t(\hat R))=\frac{1}{2}|\sin(2t\lambda)|
	\label{312b}\en
	\item if $\lambda_2\gg\lambda_1$ then
	\be
		\hat r(t)\simeq1, \qquad \hat c(t)\simeq0,\qquad \mbox{and}\qquad (\Delta_n \gamma^t(\hat R))\simeq \sqrt{\frac{\lambda_2}{\lambda_1}}|\sin(t\sqrt{\lambda_1\lambda_2})|\qquad(\Delta_n \gamma^t(\ch))\simeq0.
		\label{314b}\en
	The supporters do not perceive $\G$'s action is efficient and they are un-satisfied, but differently from the previous section the corresponding variance is not zero.
	\item if $\lambda_2\ll\lambda_1$ then
	\be
	\hat r(t)\simeq0, \qquad \hat c(t)\simeq1,\qquad \mbox{and}\qquad (\Delta_n \gamma^t(\hat R))\simeq0\qquad(\Delta_n \gamma^t(\ch))\simeq \sqrt{\frac{\lambda_1}{\lambda_2}}|\sin(t\sqrt{\lambda_1\lambda_2})|.
	\label{315b}\en
	In this case they are satisfied and their perception is positive, but with a doubt.

\end{itemize}

We can again introduce the same index of success for $\G$ as before: $\nu=\frac{\lambda_1}{\lambda_2}$. Not surprisingly, we can get the same conclusions as for $H_1$, even if $H_2$ looks different. The reason is that $H_1$ and $H_2$ really describe the same situation, but in different variables. In particular we see that if $\nu$ is large, then $\Sc$ is still very happy with what $\G$ is doing, while if $\nu\simeq0$, then $\Sc$ are very un-satisfied.

As we discussed in the previous section, we expect that here $[\gamma^t(\hat R),\gamma^t(\ch)]\neq0$. In fact, this is the case. We find:
$$
\gamma^t(\hat R)=\sin^2(t\sqrt{\lambda_1\lambda_2})\left(1-\frac{\lambda_2}{\lambda_1}\right)\hat R\hat C+\cos^2(t\sqrt{\lambda_1\lambda_2})\hat R+\frac{\lambda_2}{\lambda_1}\sin^2(t\sqrt{\lambda_1\lambda_2})\hat C+$$
\be +\frac{i}{2}\sqrt{\frac{\lambda_2}{\lambda_1}}\sin(2t\sqrt{\lambda_1\lambda_2})(c^{\dagger}r-r^{\dagger}c),
\en
and
$$
\gamma^t(\hat C)=\sin^2(t\sqrt{\lambda_1\lambda_2})\left(1-\frac{\lambda_1}{\lambda_2}\right)\hat R\hat C+\cos^2(t\sqrt{\lambda_1\lambda_2})\hat C+ \frac{\lambda_1}{\lambda_2}\sin^2(t\sqrt{\lambda_1\lambda_2})\hat R+$$
\be +\frac{i}{2}\sqrt{\frac{\lambda_1}{\lambda_2}}\sin(2t\sqrt{\lambda_1\lambda_2})(r^{\dagger}c-c^{\dagger}r),
\en
so that

\be
[\gamma^t(\hat R),\gamma^t(\ch)]=\frac{i}{2}\sin(2t\sqrt{\lambda_1\lambda_2})\left(\sqrt{\frac{\lambda_1}{\lambda_2}}-\sqrt{\frac{\lambda_2}{\lambda_1}}\right)(c^{\dagger}r+r^{\dagger}c).
\label{319b}\en
Differently from the previous section, as we expected, the commutator is in general not zero, except for some special values of $t$: $t=0$ or $t=\frac{\pi}{2\sqrt{\lambda_1\lambda_2}}$, for instance.  We can also observe that, of course, the commutator is zero when $\lambda_2=\lambda_1$. This is in agreement with our comment in Section \ref{sect2} and with the fact that, in this special case,   $H_2=H_2^\dagger$ and the corresponding time evolution operator becomes unitary. Summarizing: in the present model compatible observables became incompatible, to re-became compatible again after some time. This is really a well known (and already mentioned) phenomenon in politics: voters have no memory! After some time they just forget what $\C$ has done or has promised to do. In this perspective $\frac{\pi}{2\sqrt{\lambda_1\lambda_2}}$ can be considered as a measure of this {\em time for memory loss}.

\subsection{Third Hamiltonian}\label{sect3c}
 In the previous sections we modeled the system using Hamiltonians that describe a { rational} behavior: when the perception perceived by the electors is positive, the satisfaction increases (or analogously the lack of satisfaction decreases) and when the perception is negative, the opposite occurs. In this section we will introduce an Hamiltonian that includes two extra terms that describes an irrational behavior, possibly closer to the human {\em way of acting}.
 The Hamiltonian we want to consider now is the following:  
 \be
 H_3=\lambda_1 c^{\dagger}f^{\dagger}+\lambda_2 fc+\lambda_3c^{\dagger}f+\lambda_4f^{\dagger}c,
 \en
 which is written in terms of the same fermionic operators used in Section \ref{sect3a}.
 The last two terms in $H_3$ can be interpreted as follows: the term $c^{\dagger}f$ describes the fact that even if the perceived efficiency of $\G$ increases, the approval of $\Sc$ for $\C$ decreases. The term $cf^{\dagger}$ describes the opposite effect: even if the perceived efficiency of $\G$ decreases, the approval of $\Sc$ for $\C$ increases: these two are the irrational contributions in $H_3$. It is clear that the rational-versus-irrational behaviour of $\Sc$ is described by the relative differences between $(\lambda_1,\lambda_2)$ and $(\lambda_3,\lambda_4)$.
 
As in the previous sections, changing the values of $\lambda_i, i=1, ..., 4$, we will compute the time evolution of $\hat c(t)$ and of $\hat f(t)$, with the related variances, and comment on these results.

\vspace{2mm}

\textbf{Remark:--} The Hamiltonian $H_3$ is not balanced, if $\lambda_1\lambda_2\neq0$. Hence we do not expect some conserved quantity exists, in this model. Moreover, even rewriting the Hamiltonian in terms of $r$ and $r^{\dagger}$ as we did in $H_2$, would produce an un-balanced Hamiltonian. Hence it makes no much sense to consider a new $H_4$ with $(f,f^\dagger)$ replaced by $(r^\dagger,r)$.

\vspace{2mm}

\begin{figure}[H]
    \centering
    
    \begin{subfigure}{0.45\textwidth}
        \centering
        \includegraphics[width=\linewidth]{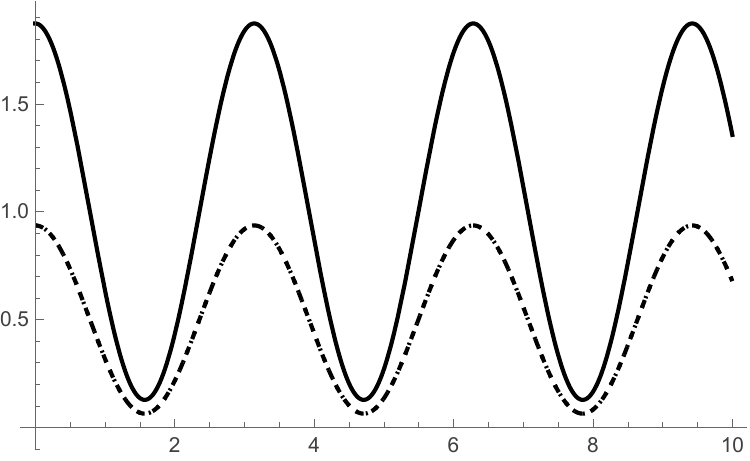}
    \end{subfigure}
    \hfill
    \begin{subfigure}{0.45\textwidth}
        \centering
        \includegraphics[width=\linewidth]{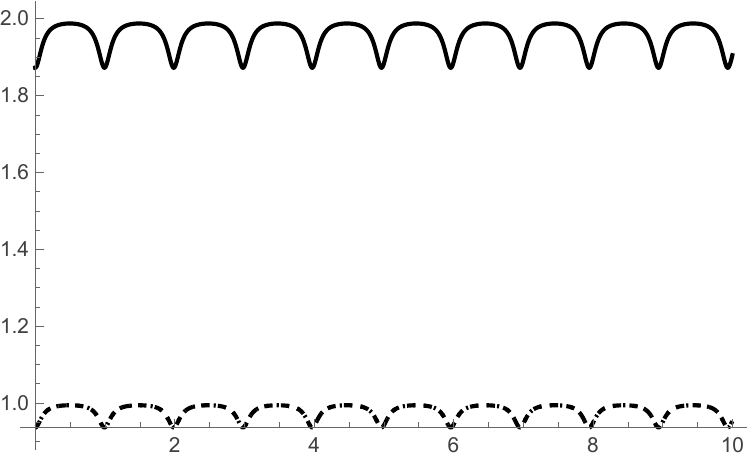}

    \end{subfigure}
    
    \vspace{0.5cm}
     \begin{subfigure}{0.45\textwidth}
        \centering
        \includegraphics[width=\linewidth]{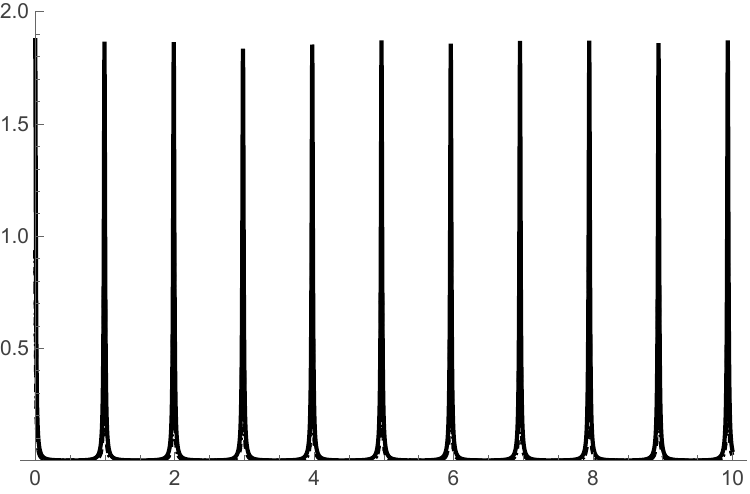}

    \end{subfigure}
    \hfill
    \begin{subfigure}{0.45\textwidth}
        \centering
        \includegraphics[width=\linewidth]{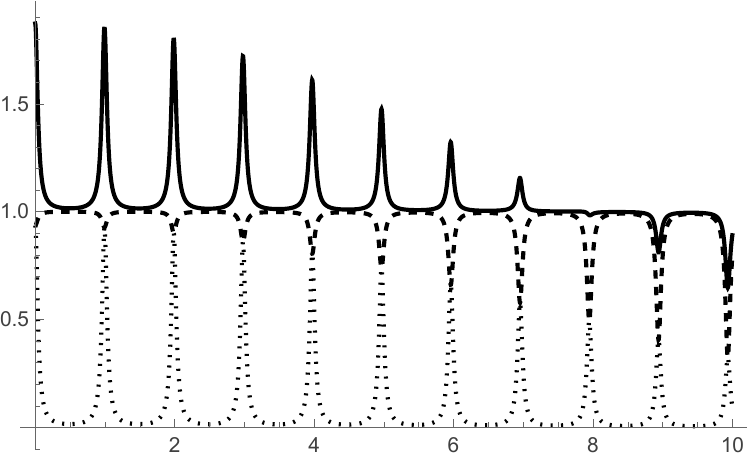}

    \end{subfigure}
    
    \caption{$\hat f(t)$ (dashed line), $\hat c(t)$ (dotted line) and their sum (thick line), for $\alpha=100$, \\$\beta=\gamma=25$ and $\eta=10$ and for $(\lambda_1, \lambda_2, \lambda_3, \lambda_4)=(1, 1, 1, 1)$ (up left), for $(\lambda_1, \lambda_2, \lambda_3, \lambda_4)=(100, 0.1, 0.1, 0.1)$ (up right), for $(\lambda_1, \lambda_2, \lambda_3, \lambda_4)=(0.1, 100, 0.1, 0.1)$ (down left) and for $(\lambda_1, \lambda_2, \lambda_3, \lambda_4)=(0.1, 0.1, 0.1, 100)$ (down right)}
\label{fig3}\end{figure}

\begin{figure}[htbp]
    \centering
    
    \begin{subfigure}{0.45\textwidth}
        \centering
        \includegraphics[width=\linewidth]{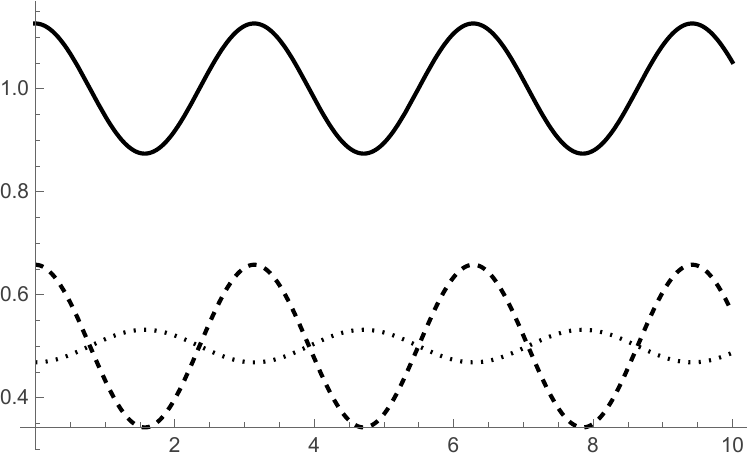}
    \end{subfigure}
    \hfill
    \begin{subfigure}{0.45\textwidth}
        \centering
        \includegraphics[width=\linewidth]{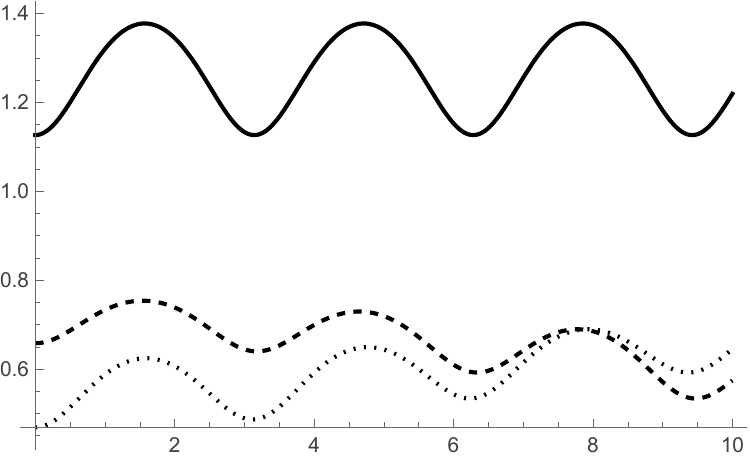}

    \end{subfigure}
    
    \vspace{0.5cm}
    
    \begin{subfigure}{0.45\textwidth}
        \centering
        \includegraphics[width=\linewidth]{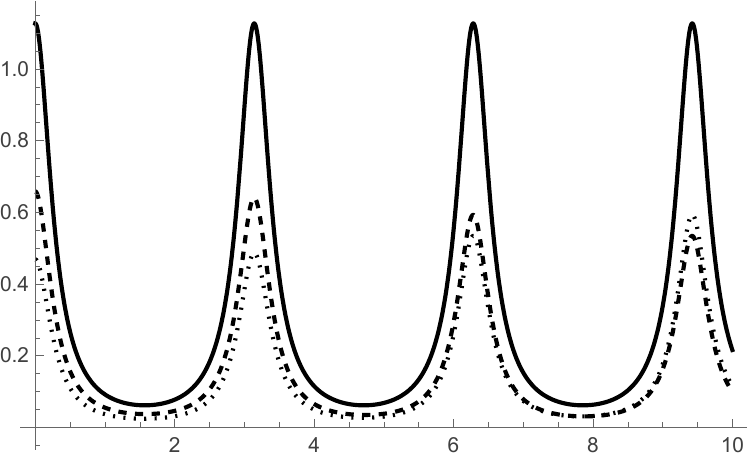}

    \end{subfigure}
    \hfill
    \begin{subfigure}{0.45\textwidth}
        \centering
        \includegraphics[width=\linewidth]{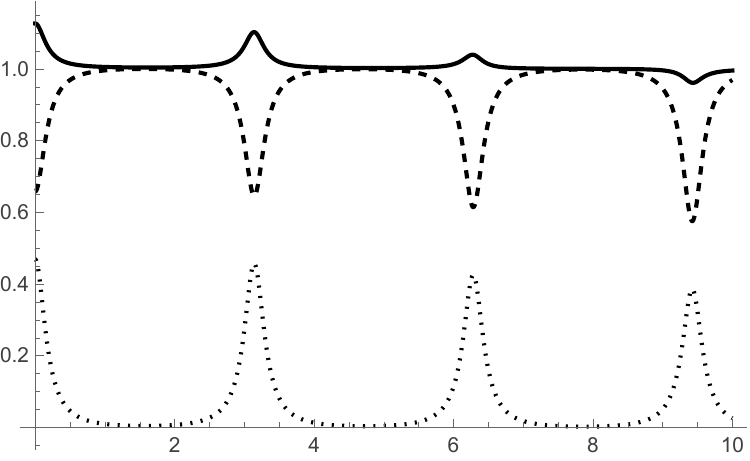}

    \end{subfigure}
    
    \caption{{$\hat f(t)$ (dashed line)}, $\hat c(t)$ (dotted line) and their sum (thick line), for $\alpha=50$, $\beta=100$, $\gamma=80$ and $\eta=10$ and for $(\lambda_1, \lambda_2, \lambda_3, \lambda_4)=(1, 1, 1, 1)$ (up left), for $(\lambda_1, \lambda_2, \lambda_3, \lambda_4)=(10, 0.1, 0.1, 0.1)$ (up right), for $(\lambda_1, \lambda_2, \lambda_3, \lambda_4)=(0.1, 10, 0.1, 0.1)$(down left) and for  $(\lambda_1, \lambda_2, \lambda_3, \lambda_4)=(0.1, 0.1, 0.1, 10)$ (down right)}
\label{fig4}\end{figure}

Before we go forward with the calculations, we introduce the initial state that will be considered in this section. This state includes not only electors who are satisfied and views the political group positively, but also electors who are weakly aligned voters.
The initial state is normalized and it is quite general:
\be\label{vecgen}
\hat \Psi(0)=\frac{\alpha\varphi_{11}+\beta\varphi_{01}+\gamma\varphi_{10}+\eta\varphi_{00}}{\sqrt{\alpha^2+\beta^2+\gamma^2+\eta^2}}, \qquad \mbox{where $\alpha, \beta, \gamma, \eta \in \mathbb{R}$.}
\en
We can interpret each contribution in (\ref{vecgen}) as in Section \ref{sect3a}:\\
$\varphi_{1,1}$ represents the electors who believe in $\C$ and in its political project. They are satisfied with the electoral proposals and their perception is positive;\\
$\varphi_{0,1}$ represents the electors whose feedback is negative even if they have a good opinion on $\G$'s (future) efficiency; \\
$\varphi_{1,0}$ represents the electors whose feedback is positive, but they have a bad opinion on $\G$'s (future) efficiency;\\
$\varphi_{0,0}$ represents the electors whose feedback is negative and have a bad opinion on $\G$'s (future) efficiency, but they decide to vote for them possibly because there are no better alternatives.\\ 
With the same strategy already adopted for $H_1$ and $H_2$, and with the same definitions for $A$ and $B$, we find that: 
\be
H_3^{2n}=(\lambda_1\lambda_2)^nA+(\lambda_3\lambda_4)^nB \qquad \mbox{and}\qquad  H_3^{2n+1}=(\lambda_1\lambda_2)^nH_1+(\lambda_3\lambda_4)^n\tilde H_2, \en
$\forall n\geq1$, and where we have introduced $\tilde H_2=\lambda_3c^{\dagger}f+\lambda_4f^{\dagger}c$, which differs from $H_2$ only because $(\lambda_1,\lambda_2)$ are now replaced by $(\lambda_3,\lambda_4)$ {and $(r, r^{\dagger})$ by $(f, f^{\dagger})$}. The time evolution operator and its adjoint are:\\
\be
e^{-itH_3}=-\frac{i}{\sqrt{\lambda_1\lambda_2}}\sin(t\sqrt{\lambda_1\lambda_2})\,H_1-\frac{i}{\sqrt{\lambda_3\lambda_4}}\sin(t\sqrt{\lambda_3\lambda_4})\,\tilde H_2+\cos(t\sqrt{\lambda_1\lambda_2})A+\cos(t\sqrt{\lambda_3\lambda_4})B,\en
and
\be
e^{itH^{\dagger}_3}=\frac{i}{\sqrt{\lambda_1\lambda_2}}\sin(t\sqrt{\lambda_1\lambda_2})\,H_1^{\dagger}+\frac{i}{\sqrt{\lambda_3\lambda_4}}\sin(t\sqrt{\lambda_3\lambda_4})\,\tilde H_2^{\dagger}+\cos(t\sqrt{\lambda_1\lambda_2})A+\cos(t\sqrt{\lambda_3\lambda_4})B.\en
Computing {$\Psi(t)=e^{-iH_3t}\hat \Psi(0)$} we get  
{
\begin{equation}
\begin{aligned}
\Psi(t) &= \frac{1}{\sqrt{\alpha^2+\beta^2+\gamma^2+\eta^2}} \Bigg[
\left({i}\sqrt{\frac{\lambda_1}{\lambda_2}}\eta
\sin(t\sqrt{\lambda_1\lambda_2})+
\alpha\cos(t\sqrt{\lambda_1\lambda_2})\right)\varphi_{11}+\\&\quad\left({-i}\sqrt{\frac{\lambda_3}{\lambda_4}}\gamma
\sin(t\sqrt{\lambda_3\lambda_4})+
\beta\cos(t\sqrt{\lambda_3\lambda_4})\right)\varphi_{01}+\\&\left({-i}\sqrt{\frac{\lambda_4}{\lambda_3}}\beta
\sin(t\sqrt{\lambda_3\lambda_4})+
\gamma\cos(t\sqrt{\lambda_3\lambda_4})\right)\varphi_{10}+\\&\quad\left({i}\sqrt{\frac{\lambda_2}{\lambda_1}}\alpha
\sin(t\sqrt{\lambda_1\lambda_2})+
\eta\cos(t\sqrt{\lambda_1\lambda_2})\right)\varphi_{00}
\Bigg],
\end{aligned}
\end{equation}
}
so that
\begin{equation}
\begin{aligned}
||\Psi(t)||^2 &=\frac{1}{\alpha^2+\beta^2+\gamma^2+\eta^2}\Bigg[\sin^2(t\sqrt{\lambda_1\lambda_2})\left(\frac{\lambda_1}{\lambda_2}\eta^2+\frac{\lambda_2}{\lambda_1}\alpha^2\right)+\sin^2(t\sqrt{\lambda_3\lambda_4})\left(\frac{\lambda_3}{\lambda_4}\gamma^2+\frac{\lambda_4}{\lambda_3}\beta^2\right)+ \\
&\quad
\cos^2(t\sqrt{\lambda_1\lambda_2})(\alpha^2+\eta^2)+\cos^2(t\sqrt{\lambda_3\lambda_4})(\beta^2+\gamma^2)\Bigg].
\end{aligned}
\end{equation}
With long but not difficult computations we find:
{\footnotesize
\be
\hat f(t)= \frac{\sin^2(t\sqrt{\lambda_1\lambda_2})\left(\frac{\lambda_1}{\lambda_2}\eta^2-\alpha^2\right)+\alpha^2+\gamma^2+\sin^2(t\sqrt{\lambda_3\lambda_4})\left(\frac{\lambda_4}{\lambda_3}\beta^2-\gamma^2\right)}{\Bigg[\sin^2(t\sqrt{\lambda_1\lambda_2})\left(\frac{\lambda_1}{\lambda_2}\eta^2+\frac{\lambda_2}{\lambda_1}\alpha^2\right)+\sin^2(t\sqrt{\lambda_3\lambda_4})\left(\frac{\lambda_3}{\lambda_4}\gamma^2+\frac{\lambda_4}{\lambda_3}\beta^2\right)+\cos^2(t\sqrt{\lambda_1\lambda_2})(\alpha^2+\eta^2)+\cos^2(t\sqrt{\lambda_3\lambda_4})(\beta^2+\gamma^2)\Bigg]}\en
}
{\footnotesize
\be
\hat c(t)= \frac{\sin^2(t\sqrt{\lambda_1\lambda_2})\left(\frac{\lambda_1}{\lambda_2}\eta^2-\alpha^2\right)+\alpha^2+\beta^2+\sin^2(t\sqrt{\lambda_3\lambda_4})\left(\frac{\lambda_3}{\lambda_4}\gamma^2-\beta^2\right)}{\Bigg[\sin^2(t\sqrt{\lambda_1\lambda_2})\left(\frac{\lambda_1}{\lambda_2}\eta^2+\frac{\lambda_2}{\lambda_1}\alpha^2\right)+\sin^2(t\sqrt{\lambda_3\lambda_4})\left(\frac{\lambda_3}{\lambda_4}\gamma^2+\frac{\lambda_4}{\lambda_3}\beta^2\right)+\cos^2(t\sqrt{\lambda_1\lambda_2})(\alpha^2+\eta^2)+\cos^2(t\sqrt{\lambda_3\lambda_4})(\beta^2+\gamma^2)\Bigg]}\en
}\normalsize
Of course, because of $H_3$ is not {\em balanced}, there is no reason to expect that $\hat f(t)+ \hat c(t)$ is conserved, except if $\lambda_1=\lambda_2=0$. In fact, with this specific choice, $H_3$ is {\em balanced}, and our general results in \cite{baglia} apply. Figure \ref{fig3} and Figure \ref{fig4} show the plots of $\hat f(t)$, $\hat c(t)$ and their sum, for different values of $\lambda$'s and different choices of $\alpha$, $\beta$, $\gamma$ and $\eta$.
All these plots represent the case in which the initial state is described by a superposition of states where the weight ($\alpha$) of voters who are convinced in favor of $\C$ is the largest, while the weights ($\beta$, $\gamma$ and $\eta$) of voters who vote mainly because they lack better alternatives are smaller. We chose this initial state because it can represent a good realistic scenario: in a real group of voters, a big part of them votes because they are convinced, some because they are satisfied with the candidate’s proposals or perceive them as efficient, and few others because they have no better alternatives. In the figure up left the system is driven by an Hamiltonian in which the values of $\lambda$ are the same and, for the choice of the coefficients of the initial state, $\hat f (t)$ and $\hat c(t)$ behave in the same (purely periodic) way, as expected. In the other figures, we consider a single value of $\lambda$ significantly larger than the others to explore what this a-symmetry implies in the time evolution of the (decision) functions plotted.\\
In Figure 4, we chose values of $\beta$ and $\gamma$ larger than the others in order to represent an initial state in which the electors vote for a political group solely because they appreciate its political proposals or perceive the group as efficient while being in the political minority.

The computation of the variances is long and the results strongly depend on the parameters of the model, i.e. on the $\lambda$'s and on the parameters defining $\hat\Psi(0)$ in (\ref{vecgen}) . For instance, if we assume that $\alpha>>\beta, \gamma, \eta$, we get
{\footnotesize$$
\Delta_n\gamma^t(\hat X)=\sqrt{\left|\frac{\cos^2(t\sqrt{\lambda_1\lambda_2})}{\frac{\lambda_2}{\lambda_1}\sin^2(t\sqrt{\lambda_1\lambda_2})+\cos^2(t\sqrt{\lambda_1\lambda_2})}\Big[\frac{\lambda_1}{\lambda_2}\sin^2(t\sqrt{\lambda_1\lambda_2})+\cos^2(t\sqrt{\lambda_1\lambda_2})-\frac{\cos^2(t\sqrt{\lambda_1\lambda_2})}{\frac{\lambda_2}{\lambda_1}\sin^2(t\sqrt{\lambda_1\lambda_2})+\cos^2(t\sqrt{\lambda_1\lambda_2})}\Big]\right|},
$$}
both for $\hat X=\hat C$ and for $\hat X=\hat F$. Notice that, if  $\lambda_1=\lambda_2=:\lambda$ they reduce to:
$$\Delta_n\gamma^t(\hat F)=\Delta_n\gamma^t(\hat C)=\frac{1}{2}|\sin(2t\lambda)|.$$
Moreover we see that, for $\lambda_1>>\lambda_2$,
$$\Delta_n\gamma^t(\hat F)=\Delta_n\gamma^t(\hat C)=\sqrt{\left|\frac{\lambda_1}{\lambda_2}\sin^2(t\sqrt{\lambda_1\lambda_2})-1\right|},$$
while for $\lambda_1<<\lambda_2$ we find
$$\Delta_n\gamma^t(\hat F)=\Delta_n\gamma^t(\hat C)\simeq0.$$
Similar results can be deduced with other choices for $\alpha$, $\beta$, $\gamma$ and $\eta$, but these are not particularly interesting and will not be given here.

The analysis of $\hat f(t)$, $\hat c(t)$ and their related variances is very rich here. For instance, in order to have $\hat f(t)=0$ for some $t$, we should have, for instance, $t=\frac{k\pi}{\sqrt{\lambda_1\lambda_2}}=\frac{l\pi}{\sqrt{\lambda_3\lambda_4}}$, for some integer $k,l$ and $\alpha=\gamma=0$. This is, clearly, only possible for specific values of the $\lambda_j$'s. It is also clear that other possibilities also exist.

We are more interested in the compatibility of the observables. For that, we first need to compute the time evolution of $\hat F$ and $\hat C$, which are the following:
\be
\begin{aligned}
\gamma^t(\hat F)&= \hat C\left(-\frac{\lambda_1}{\lambda_2}\sin^2(t\sqrt{\lambda_1\lambda_2})+\frac{\lambda_4}{\lambda_3}\sin^2(t\sqrt{\lambda_3\lambda_4})\right)+\hat F\left(-\frac{\lambda_1}{\lambda_2}\sin^2(t\sqrt{\lambda_1\lambda_2})+\cos^2(t\sqrt{\lambda_3\lambda_4})\right)\\ &\quad +\hat F \hat C\left(\frac{\lambda_1}{\lambda_2}\sin^2(t\sqrt{\lambda_1\lambda_2})+\cos^2(t\sqrt{\lambda_1\lambda_2})-\frac{\lambda_4}{\lambda_3}\sin^2(t\sqrt{\lambda_3\lambda_4})-\cos^2(t\sqrt{\lambda_3\lambda_4})\right)\\&\quad+\frac{i}{2}\sqrt{\frac{\lambda_1}{\lambda_2}}\sin(2t\sqrt{\lambda_1\lambda_2})(f^{\dagger}c^{\dagger}+fc)+\frac{i}{2}\sqrt{\frac{\lambda_4}{\lambda_3}}\sin(2t\sqrt{\lambda_3\lambda_4})(c^{\dagger}f-f^{\dagger}c)+\frac{\lambda_1}{\lambda_2}\sin^2(t\sqrt{\lambda_1\lambda_2})\1
\end{aligned}
\en
and
\be
\begin{aligned}
\gamma^t(\hat C)&= \hat C\left(-\frac{\lambda_1}{\lambda_2}\sin^2(t\sqrt{\lambda_1\lambda_2})+\cos^2(t\sqrt{\lambda_3\lambda_4})\right)+\hat F\left(-\frac{\lambda_1}{\lambda_2}\sin^2(t\sqrt{\lambda_1\lambda_2})+\frac{\lambda_3}{\lambda_4}\sin^2(t\sqrt{\lambda_3\lambda_4})\right)\\ &\quad +\hat F \hat C\left(\frac{\lambda_1}{\lambda_2}\sin^2(t\sqrt{\lambda_1\lambda_2})+\cos^2(t\sqrt{\lambda_1\lambda_2})-\frac{\lambda_3}{\lambda_4}\sin^2(t\sqrt{\lambda_3\lambda_4})-\cos^2(t\sqrt{\lambda_3\lambda_4})\right)\\&\quad+\frac{i}{2}\sqrt{\frac{\lambda_1}{\lambda_2}}\sin(2t\sqrt{\lambda_1\lambda_2})(fc-c^{\dagger}f^{\dagger})+\frac{i}{2}\sqrt{\frac{\lambda_3}{\lambda_4}}\sin(2t\sqrt{\lambda_3\lambda_4})(f^{\dagger}c-c^{\dagger}f)+\frac{\lambda_1}{\lambda_2}\sin^2(t\sqrt{\lambda_1\lambda_2})\1
\end{aligned}
\en
Notice that, indeed, we have $\gamma^0(\hat C)=\hat C$ and $\gamma^0(\hat F)=\hat F$, as it must be.

From these formulas we get: 
\be
[\gamma^t(\hat F),\gamma^t(\hat C)]=\frac{i}{2}\sin(2t\sqrt{\lambda_3\lambda_4})\Big(\sqrt{\frac{\lambda_3}{\lambda_4}}-\sqrt{\frac{\lambda_4}{\lambda_3}}\Big)(c^{\dagger}f+f^{\dagger}c)
\en
Incidentally we observe that this formula returns the original commutativity between $\hat C$ and $\hat F$: $[\gamma^0(\hat C),\gamma^0(\hat F)]=0$, as expected. We observe that this quantity is, in general, non zero and we can also observe that this result is similar to the one deduced in the previous section. Indeed, the terms that give a commutator different from zero are the terms that do not increase and decrease the eigenvalues of $\hat F$ and $\ch$ together. The commutator is zero for some special values of $t$, so { incompatible} observables become at some time { compatible}, and it becomes identically zero when $\lambda_3=\lambda_4$. This is in agreement with our conclusions in Section \ref{sect3a}, where the role of $\lambda_{1,2}$ was inessential in the computation of $[\gamma^t(\hat F),\gamma^t(\hat C)]$.

\section{Some comments on the dynamics}\label{sectconfronto}

Quantum-like ideas have been proposed in the past few decades in very different applications, from economics to biology, from decision making to ecology. In many cases, the relevant aspect of the analysis is the dynamical behavior of the system under consideration, which is already a very important aspect in ordinary quantum mechanics, and a very rich topic of research. The Schr\"odinger equation for the wave function, the master equation for the density matrix, the Lindblad approach to open systems, are just some of the possible ways to deduce the time evolution of a given system, \cite{mess,mer,ali,benatti}. Other approaches also exist: a purely (self-adjoint) Hamiltonian approach for open systems originally used in quantum optics, \cite{barrad}, and later adopted for macroscopic systems, see \cite{bbk} for instance. Or the so-called $(H,\rho)$-induced dynamics, \cite{bdsgo}, which was proposed to model systems with some dynamics {\em adjusting itself} because of some external factor. Or yet, in a possibly more pragmatic way and in connection with applications outside the quantum world, simple master equations have been adopted, \cite{asano1,asano2,asano3}. All these approaches
have, not surprisingly, pros and contra. A common pro, which has been analyzed in many details by one of us (F.B.) is the existence of some {\em asymptotic limit} for some specific functions, the so-called {\em decision functions}. Since they describe the procedure of taking a decision, it is clear that one would like these functions to converge to some value, which describes exactly  the decision taken by the agent whose behavior we want to deduce. This feature can be described in some complicated, but very detailed ways, using for instance an Hamiltonian approach for systems with infinite degrees of freedom, \cite{bbk}, or with a much simpler master equation, \cite{asano2,asano3}, or using the $(H,\rho)$-induced dynamics, \cite{bdsgo}. Another possibility is to work as we did here, i.e. introducing some non self-adjoint Hamiltonian as the {\em generator} of the dynamics. In particular, reaching some equilibrium and proving the existence of conserved quantities is possible if we adopt balanced Hamiltonians, \cite{baglia}, as we have widely discussed in Section \ref{sect3}. The general result can be found in Proposition \ref{prop} in the Appendix. From this point of view, therefore, all the possibilities we have mentioned in this section are equivalent, at least for what concerns the result: they all produce asymptotic values\footnote{Except for the Schr\"odinger equation, in general.}. However, they are very different from the point of view of the technical aspects. 

Moreover, in our opinion, among all the possibilities, the $\gamma$-dynamics is the simplest, and most natural, procedure allowing a change in the nature of the observables with time: from compatible to incompatible, or viceversa. The reason is that $\gamma^t$ is not an automorphism, while the other approaches produce time evolution which are in general automorphisms: the time evolution of a product of observables is the product of the time evolution. For this reason we believe that our approach is rather efficient in this kind of problems, and in decision making in particular.

\section{Conclusions}\label{sectconcl}

In this paper we have adopted a quantum-like approach to describe a system of political parties $\C$ just winning some political elections and their supporters $\Sc$. We have shown how, using an Hamiltonian approach, it is possible to describe a sort of {\em degree of satisfaction} of  $\Sc$ in connection with what $\G$, the government formed by $\C$, is doing. We have compared the effect of three different Hamiltonians, balanced or not, and we have deduced the time evolution of some mean values representing the perceived efficiency of $\G$ and the satisfaction of $\Sc$. We have also computed the various variances, which have been interpreted here as a sort of {\em perplexity} of $\Sc$: the smaller their values, the sharper $\Sc$'s opinion on $\G$.

We have further analyzed an aspect of the system which, in our opinion, is quite interesting in several concrete situation, and, in particular, in Decision Making. This is the dynamical behavior of compatible observables. In particular we have shown, with specific examples, that if we adopt some non-Hermitian Hamiltonian to describe the time evolution of a given system, it might likely happen that two questions (or, in general, observables) which are compatible at time $t=0$, become incompatible at larger time, but they can turn back to be compatible again after still other time. This is an interesting {\em lack of memory effect}, which is quite often observed in politics, and in Decision Making in general.

Of course we could extend this research in different ways: with other Hamiltonians, with different (physical, social, political, biological,...) systems, using different ladder operators, or adopting a Lindblad (rather than a purely Hamiltonian) approach. We plan to consider some of these points soon. In our plans we would also like to relate our results to some concrete election poll.

\section*{Acknowledgements}

 F. B.  acknowledges partial  support from Palermo University and from G.N.F.M. of the INdAM. 

\section*{Fundings}

This work didn't received any fundings.

\section*{Conflicts of interest}

There are no conflicts of interest.

\section*{Availability of data and material}

Not applicable.

\section*{Code availability}

Not applicable.

\section*{Authors' contributions}

F.B. is responsible of the idea, the computations and writing. G.L. contributes in the computations and in writing the paper.

\renewcommand{\theequation}{A.\arabic{equation}}

\section*{Appendix: The $\gamma$-dynamics}\label{appendix}

Given a time-independent $H\in\B(\Hil)$ with $H\neq H^\dagger$, and $X\in\B(\Hil)$ we define the $\gamma$-dynamics as follows:	
\be
\gamma^t(X)=e^{iH^\dagger t}Xe^{-iHt},
\label{a1}\en
and the related $\gamma$-derivation as
\be
\delta_\gamma(X)=\|.\|-\lim_{t,0}\frac{\gamma^t(X)-X}{t}=i\left(H^\dagger X-XH\right).
\label{a2}\en
Notice that $(\gamma^t(X))^\dagger=\gamma^t(X^\dagger)$, for all $X\in\B(\Hil)$. Among the other results we know that the following statements are equivalent: 1) $\delta_\gamma$ is a *-derivation; 2) $\delta_\gamma(\1)=0$; 3) $H=H^\dagger$; 4) $\gamma^t(\1)=\1$; 5)  $\gamma^t(XY)=\gamma^t(X)\gamma^t(Y)$, $\forall X,Y\in\B(\Hil)$. This implies that, whenever $H\neq H^\dagger$, none of the other properties listed here holds true.
	
Taken  $X\in\B(\Hil)$, the following statements are  equivalent: 1) $H^\dagger X=XH$; 2) $\delta_\gamma(X)=0$; 3) $ \gamma^t(X)=X$. When $X$ satisfies these conditions, $X$ is called a $\gamma$-symmetry.
	
In \cite{baglia} we have discussed the importance of taking the mean values of the relevant observables on vectors which are normalized at each time. More explicitly, let us assume that the wave function of the system $\Psi(t)$ obeys the Schr\"odinger equation $i\dot\Psi(t)=H\Psi(t)$, where (in general) $H\neq H^\dagger$. Hence we can write $\Psi(t)=e^{-iHt}\Psi(0)$, where $\Psi(0)$ describes the original state of the system. Suppose $X\in\B(\Hil)$ is an observable describing some specific feature of the system which we want to compute when $t$ changes. Hence, see \cite{bagbook,bagbook2,FFF} and references therein, we could compute 
\be
x(t)=\pin{\Psi(0)}{\gamma^t(X)\Psi(0)}=\pin{\Psi(t)}{X\Psi(t)}.\label{a3}\en 
However, if $H\neq H^\dagger$, $x(t)$ could assume {\em strange} values, \cite{baglia}. For this reason, it is more reasonable to compute a normalized version of $x(t)$:
\be
\hat x(t)=\pin{\hat\Psi(t)}{X\hat\Psi(t)}, \qquad \mbox{where}\qquad \hat\Psi(t)=\frac{\Psi(t)}{\norm{\Psi(t)}}.
\label{a4}\en
It is possible to see that $\hat\Psi(t)$ obeys a sort of nonlinear Schr\"odinger equation, \cite{baglia}, and that (at least for finite dimensional systems) $\|\Psi(t)\|\neq0$ for all $t$. In this same paper we have proved an interesting result which gives sufficient conditions for $\hat x(t)$ to stay constant in time: let us suppose that the Hamiltonian $H$ of the system, Hermitian or not, can be written in terms of some ladder operators $b_j$ and $b_j^\dagger$, $j=1,2,\ldots,L$, not necessarily of the same nature, i.e. not necessarily obeying the same algebraic rules. Assume further that $H=\sum_{k=1}^MH_k$, and that each $H_K$ is the product of different $b_j$ and $b_j^\dagger$. If the number of lowering and raising operators in each $H_k$ coincides, we say that $H$ is {\em balanced}. Let us now further introduce $\hat N_{tot}=\sum_{j=1}^L\hat n_j$, $\hat n_j=b_j^\dagger b_j$, the {\em total number} operator for $\Sc$. Then the following result holds:

\begin{prop}\label{prop}
	Under the above assumptions, if $\Psi(0)$ is an eigenstate of all the $\hat n_j$, it follows that $\pin{\hat\Psi(t)}{X\hat\Psi(t)}$ is constant in time.
\end{prop}

The proof of this proposition, together with several examples and comments, can be found in \cite{baglia}. A specific application of this Proposition is what we have deduced in Section \ref{sect3b}. It is worth stressing that, \cite{baglia}, balanced Hamiltonians do not necessarily give rise to conserved quantities, if the assumption on $\Psi(0)$ is not satisfied.

\end{document}